# Self-formation of coherent emission in a cavity-free system

A. A. Zyablovsky[1,2], I. V. Doronin[1,2], E. S. Andrianov[1,2], A. A. Pukhov[1,2,3], Yu. E. Lozovik[1,2,4], A. P. Vinogradov[1,2,3], and A. A. Lisyansky[5,6]

[1]Dukhov Research Institute of Automatics (VNIIA), 127055, 22 Sushchevskaya, Moscow, Russia

[2]Moscow Institute of Physics and Technology, 141700, 9 Institutskiy per., Moscow, Russia

[3]Institute for Theoretical and Applied Electromagnetics, 125412, 13 Izhorskaya, Moscow, Russia

[4]Institute of Spectroscopy Russian Academy of Sciences, 108840, 5 Fizicheskaya, Troitsk, Moscow, Russia

[5]Department of Physics, Queens College of the City University of New York, Queens, New York 11367, USA

[6]The Graduate Center of the City University of New York, New York, 10016, USA

It is commonly accepted that a collection of pumped atoms without a resonator, which provides feedback, cannot lase. We show that intermodal coupling via active atoms pulls the frequencies of the free-space modes towards the transition frequency of the atoms. Although at a low pump rate mode phases randomly fluctuate, phase realizations at which interference of pulled modes is constructive emerge. This results in an increase of stimulated emission into such realizations and makes their lifetime longer. Thus, mode pulling provides positive feedback. When the pump rate exceeds a certain threshold, the lifetime of one of the realizations diverges, and radiation becomes coherent.

*Introduction.* – Radiation of a collection of incoherently pumped atoms in a cavity-free system is usually referred to as amplified spontaneous emission (ASE) or superfluorescence [1]. It takes place in various physical systems – from so-called cosmic lasers to superluminescent diodes [2-8]. In such a system, a photon spontaneously emitted by one of the excited atoms triggers stimulated emission of inverted atoms located on the path out of the system [1]. Thus, ASE is a result of an interplay of stimulated and spontaneous emission, in contrast to lasing that occurs due to stimulated emission. Nevertheless, both phenomena have some common features.

First, in both laser and ASE systems, line narrowing of output radiation is observed. Second, similar to lasing, the ASE intensity output has an *S*-shaped dependence on the pump rate. These features arise due to nonlinear amplification of radiation. In lasers, a resonator turns an amplifier into a generator making radiation coherent. It is commonly believed that a cavity-free system cannot generate coherent radiation. In other words, an ASE system can only serve as an incoherent light source or as a light amplifier (see for detail Ref. [9]).

In this paper, we show that even without a cavity, there is a mechanism for creating positive feedback, in ASE systems. This mechanism arises due to the interaction of free space modes with pumped atoms. This interaction changes the density of states (DOS) of relevant modes of the environment.

We consider a collection of pumped atoms surrounded by free space. The free space is modeled by a one-dimensional multi-mode optical waveguide. By computer simulation, we demonstrate that free-space modes indirectly interact via inverted atoms. This interaction is responsible for frequency pulling, in which frequencies of free space modes are attracted to the transition frequency of the atoms of the active medium. This results in a maximum of the DOS at the transition frequency. Although near the maximum, the mode frequencies are close, their phases fluctuate because these modes are supported by spontaneous emission. Nevertheless, some sets of phases that lead to constructive interference in the modes emerge. In such a fluctuation, the modes take away the energy from the pumped atoms more efficiently, and the fluctuation has a greater lifetime. We find the parameters of a cavity-free system for which the lifetime of such fluctuations may diverge, and the stimulated emission of the inverted atoms results in self-oscillations (lasing).

*The model and main equations.* – In a typical experiment on ASE, a large volume is usually filled with active atoms. Only a fraction of these atoms within a region stretched in one direction is pumped. Assuming that the transverse dimensions of the volume are much smaller than the longitudinal, we, for simplicity, consider a one-dimensional multimode waveguide extended in the *x*-direction. To make this model closer to a real ASE experiment, we take into account the radiation losses in the 1*D* waveguide walls that correspond to losses through side boundaries of the stretched 3*D* region.

In order to quantize the electromagnetic (EM) field, we first consider a finite system, bounded by ideally reflecting walls separated by a very large distance $L_U$ acting as the size of the universe. We then let $L_U$ (and consequently the round-trip time $t_U = L_U / c$) tend to infinity.

We expand the field operator over the basis of creation and annihilation operators of the waveguides modes, $\hat{\mathbf{E}}(x) = \sum_n \mathbf{e}_n \sqrt{4\pi\hbar\omega_n / (L_U A)} \exp(i k_n x) \hat{a}_n + h.c.$, where $\hat{a}_n^\dagger$ and $\hat{a}_n$ are the creation and annihilation operators for the *n*-th free-space mode of the EM field with the wavenumber $k_n = \pi n / L_U$ and the polarization vector $\mathbf{e}_n$, $A$ is the waveguide cross-section. [See for details Sec. 1 of the Supplementary Materials (SM)].

The pumped atoms are considered as two-level systems (TLSs), occupying a finite interval, $L_{am}$, of the 1*D*-waveguide. To describe the interaction between these atoms and EM modes of the 1D-universe, we use the Heisenberg-Langevin equation with the Jaynes-Cummings Hamiltonian in the rotating wave approximation [10,11]:

$$\hat{H} = \sum_n \hbar\omega_n \hat{a}_n^\dagger \hat{a}_n + \sum_j \hbar\omega_{TLS} \hat{\sigma}_j^\dagger \hat{\sigma}_j + \sum_{n,j} \hbar\Omega_{nj}\left(\hat{a}_n^\dagger \hat{\sigma}_j + \hat{a}_n \hat{\sigma}_j^\dagger\right), \quad (1)$$

where $\omega_{TLS}$ and $\lambda_{TLS} = 2\pi c/\omega_{TLS}$ are the transition frequency and the corresponding wavelength of the two-level atoms, and the operators $\hat{\sigma}_j^\dagger$ and $\hat{\sigma}_j$ are the raising and lowering operators for the $j$-th atom of the active medium. The third term in Eq. (1) describes the dipole interaction between field modes and dipole moments of TLSs. The coupling constant $\Omega_{nj}$ (the Rabi frequency) is $\hbar\Omega_{nj} = -\mathbf{E}_n(x_j)\cdot\mathbf{d}_j = -\mathbf{d}_j \cdot \sqrt{4\pi\hbar\omega_n/(L_U A)}\left(\mathbf{e}_n \hat{a}_n \exp(ik_n x) + h.c.\right)$ and $x_j$ is the position of the $j$-th TLS with the dipole moment matrix element $\mathbf{d}_j$ (see Sec. 1 of the SM). Our estimations show that the reflection of waves at the boundary of the ensemble is so small that it does not affect the spatial distribution of the field (see Sec. 3 of the SM). Thus, the system may be considered as a cavity-free system. Since the atoms occupy a finite space interval, in the limit $L_U \to \infty$, the geometry of the system becomes typical for ASE.

In most ASE experiments, even in a subwavelength volume, the number of atoms is very large. In our consideration, to limit the number of equations, non-inverted atoms outside of the stretched region are replaced with vacuum. Such a replacement does not increase reflection at the boundaries of the active medium (see Sec. 3 of the SM) and cannot cause lasing associated with this reflection. Losses in the removed non-inverted atoms are modeled with losses in the waveguide walls.

Even though only a fraction of all atoms is pumped, their number is still large. This allows us to divide the whole volume into cells of the size $\lambda_{TLS}/10$ and consider operators averaged over the cells. Following the method of the system size expansion [12], we consider the inverse number of atoms in a cell, $1/N_c$, as an expansion parameter, and investigate the behavior of the system in the limit $N_c \gg 1$, neglecting higher in $1/N_c$ order terms. In this limit, the expected values of operators grow with $N_c$ faster than quantum corrections to these expected values [12]. Therefore, in the leading of $1/N_c$ order, the system can be described by equations for c-numbers, while in the second order of $1/N_c$, quantum corrections can be presented as a classical noise [13]. These equations are given below and are derived in Sec. 1 of the SM:

$$da_n/dt = \left(-\gamma_a/2 - i(\omega_n - \omega_{TLS})\right)a_n - i\sum_k \Omega_{nk} N_c \sigma_k^{cell}, \quad (2)$$

$$d\sigma_k^{cell}/dt = -\sigma_k^{cell}(\gamma_P + \gamma_D + \gamma_{deph})/2 + i\sum_n \Omega_{nk} a_n D_k^{cell} + F_k^{\sigma\,cell}, \quad (3)$$

$$dD_k^{cell}/dt = -(\gamma_P + \gamma_D)D_k^{cell} + (\gamma_P - \gamma_D) + 2i\sum_n \Omega_{nk}\left(a_n^* \sigma_k^{cell} - a_n\left(\sigma_k^{cell}\right)^*\right), \quad (4)$$

where $\gamma_a$ describes losses in the waveguide walls, $a_n$ is the slowly varying amplitude of the *n*-th mode and the variables $\sigma_k^{cell}$ and $D_k^{cell}$ are expected values of the lowering operators for the atoms of the active medium and the operator of the population inversion, averaged over the *k*-th cell, respectively. The coupling constant $\Omega_{nk}$ between the *n*-th mode and the averaged dipole moment of the *k*-th cell is equal to $\Omega_{nk} = -\mathbf{d}_{egk} \cdot \mathbf{E}_n(x_k)/\hbar$, where $x_k$ is the coordinate of the *k*-th cell, and $\mathbf{d}_{egk}$ is the matrix element of the transition for the dipole moment of atoms averaged over the *k*-th cell. $\gamma_D$ and $\gamma_{deph}$ describe energy losses in the active medium and the phase relaxation of the dipole moment operator, respectively, and $\gamma_P$ is the rate of incoherent pumping. The last term, $F_k^{\sigma\, cell}$, in Eq. (3) describes classical $\delta$-correlated noises (their correlation functions are given in Sec. 1 of the SM). The existence of these noise terms is essential for the description of ASE because the latter arises due to spontaneous emission of inverted atoms. Indeed, unlike stimulated radiation, which can be considered as a forced evolution of an excited atom under the action of a classical field, spontaneous radiation is a quantum phenomenon associated with an exit of an atom from a stationary excited state without external influence. In order to describe this probabilistic process in the framework of regular equations, it is necessary to introduce noise [10]. The second random process included in $F_k^{\sigma\, cell}$ is connected with incoherent pumping. In Eq. (3), both processes are described simultaneously via the interaction with a pump reservoir with negative temperature. Averaging over the variables of this reservoir brings the terms proportional to the pump rate $\gamma_p$ and the noise terms into equations.

When solving Eq. (2)-(4), we use the values of parameters that are close to those of a gain medium based on organic semiconductors. The TLS transition frequency is $\omega_{TLS} \sim 10^{15}$ rad/s, the relaxation rates are $\gamma_a = 4 \times 10^{-3} \omega_{TLS}$, $\gamma_{deph} = 10^{-2} \omega_{TLS}$, and $\gamma_D = 10^{-6} \omega_{TLS}$, the pump rate varies within the interval $(0, 1000\gamma_D)$. Since the interaction of modes with atoms is resonant, we consider a finite frequency interval $(\omega_{TLS} - 20\gamma_\sigma, \omega_{TLS} + 20\gamma_\sigma)$. The value of $L_U$ is chosen large enough to provide about 1600 equidistant modes within the interval.

In our computer simulation, we hold $L_U \gamma_a / c \gg 1$. This condition guarantees that the EM signal that leaves the active medium and then reflected from the boundaries of the universe does not return back. If the condition $L_U \gamma_a / c \gg 1$ is not satisfied, then the system can be considered as cavity-free, only for the simulation time, $t_s$, smaller than $(L_U - L_{am})/c$. Our computer simulation shows that for $t_s < (L_U - L_{am})/c$ the results obtained for the system with $L_U \gamma_a / c \leq 1$ qualitatively agree with the results obtained for the cavity-free system.

*Computer simulation of the ASE regime.* – We use Eqs. (2)-(4) to find the intensity, the spectrum, and the second-order coherence function $g_2(0)$ of the output radiation for times $t \gg \gamma_D^{-1}, \gamma_a^{-1}, \gamma_{deph}^{-1}, \gamma_P^{-1}$.

First, we make sure that the system of Eqs. (2)-(4) can describe a typical behavior of an ASE (see, e.g. [7,8]). For the system with the length of the active medium $L_{am} = 35\lambda_{TLS}$ and $G(N_c)L_{am} = 5$, where $G$ is the gain coefficient in the active medium (the number of atoms in a cell is $N_c = 1500$), our computer simulation demonstrates that below some pump rate $\gamma_P = \gamma_{comp}$, losses are greater than the energy pumped into the system and the amplitudes of the waves transmitted through the active medium diminish. Consequently, the intensity of the EM field reaches a maximum at the center of the active region [1]. Above $\gamma_{comp}$, the transmitted waves are intensified, and the maxima of the EM field appear near the edges of the active region [1]. Above the compensation threshold, $\gamma_P > \gamma_{comp}$, the dependence of the intensity of the output EM field on the pump rate has an *S*-shape [1]. The inflection point of the curve is usually considered as the ASE-threshold, $\gamma_{ASE}$ [1]. This behavior is characteristic for ASE.

To characterize coherent properties of light, we employ the second-order coherence function, $g^{(2)}(\tau) = \langle I(t)I(t+\tau)\rangle / \langle I(t)\rangle^2$. For incoherent light of a black body, $g^{(2)}(0) = 2$, while for lasers, $g^{(2)}(0) = 1$ [11]. Since in experiment, to measure $g^{(2)}(0)$, a spectrum of the investigated source is narrowed by filtering [11], we calculate $g^{(2)}(\omega_{TLS}, 0) = \langle I_{\omega_{TLS}}(t) I_{\omega_{TLS}}(t)\rangle / \langle I_{\omega_{TLS}}(t)\rangle^2$ in a narrow frequency interval near the transition frequency of active atoms. We confirm that for ASE, the second-order coherence function $g^{(2)}(\omega_{TLS}, 0)$ does not depend on the pump rate $\gamma_P$ and is about 2 both below and above the threshold $\gamma_{ASE}$ (see Ref. [9]). Thus, for the range of parameters considered above, the results of our computer simulation agree with experiments [7,8].

For $G(N_c)L_{am} > 5$, our model, however, demonstrates an unexpected behavior of cavity-free ASE systems with an increase in the atom concentration. For $N_c > 1600$, we find that there exists a new pumping threshold $\gamma_{coh} > \gamma_{ASE}$, above which the value of $g^{(2)}(\omega_{TLS}, 0)$ drops to unity (see also Secs. 4 and 5 in the SM), which is typical for lasers (Fig. 1) [9].

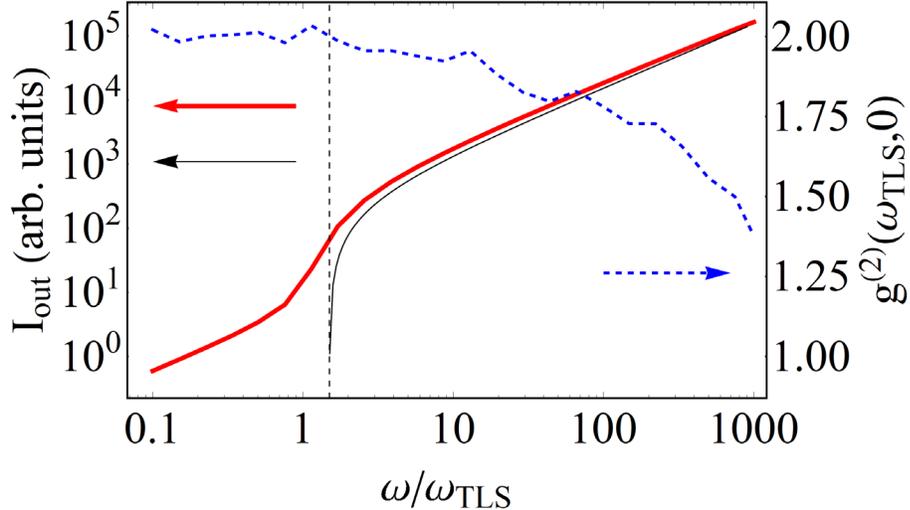

Fig. 1. The dependence of the intensity of output radiation and $g^{(2)}(\omega_{TLS},0)$ on the pump rate for the cavity-free system exhibiting lasing. The solid thick and the solid thin curves are output intensities obtained by solving the Maxwell-Bloch equation with and without noise, respectively. The dashed blue curve is the second-order correlation function, $g^{(2)}(\omega_{TLS},0)$. The vertical dashed line shows the coherence threshold $\gamma_{coh}$.

We have specially monitored the possibility of lasing due to the formation of the Fabry-Perot resonator. Our evaluation shows that in such a resonator, the lasing can never be achieved (see Secs. 2 and 3 of the SM). Moreover, as we show in Secs. 4 and 5 of the SM, the lasing arises even in a system having only a single inverted atom. Obviously, in this limiting case, no cavity can be formed.

*Mechanism for the cavity free lasing.* – As discussed in the previous section, our model demonstrates an unexpected result – lasing in a cavity-free ASE system. To understand this effect, it is useful to take a look at the textbook picture of lasing in a system with a resonator from a nonconventional point of view. An open resonator placed in free space leads to a local maximum in the DOS at the resonance frequency [14]. The modes forming the maximum have almost the same frequencies and can interfere. A certain configuration of these modes interfere constructively increasing the field intensity. Consequently, the excited atoms are stimulated to emit photons into the configuration of the modes more intensively, increasing the field further. Thus, the DOS maximum serves as positive feedback providing coherent radiation of the system. The lasing is attributed to the DOS maximum generated by a resonator [14].

Below we show that the mechanism, in which a DOS maximum provides positive feedback, can be realized in a cavity-free system. The main difference between our mechanism and the mechanism described in Ref. [14] is that in a cavity-free system, the DOS maximum is created by the nonlinear interaction of free-space modes with an active medium.

The analysis shows that without noise new eigensolutions to Eqs. (2)–(4), hybrids of free space modes and atomic states are formed. When the pump rate is near zero, these modes differ only slightly from the free-space modes, but they have complex frequencies. None of these modes can be obtained as a linear combination of free space modes because each of them has its own frequency but is described by the whole set of wave vectors of free space modes. As the pump rate increases, little happens to almost all modes. They change slightly in shape and remain strongly damped. However, for one of the modes, its damping decreases to zero, and the spatial distribution tends to a typical field distribution in an ASE system. The field intensity of the solution increases approaching the boundaries of the active medium from within. Outside the region filled with the active medium, the intensity diminishes due to losses in the waveguide walls. The field intensity has a minimum in the middle and maxima at the edges of the active region. The frequency of this mode is equal to the atom transition frequency. The phases of the spatial harmonics, forming such a distribution of the field, are fixed. When the pump threshold is reached, the imaginary part of the frequency turns to zero, and a Hopf bifurcation arises. A non-zero solution is a self-oscillation with a transition frequency of the active medium. The dependence of the intensity of the state on the pump rate is shown in Fig. 1 by the red line.

For different pump rates, the distribution of the EM field in this mode is shown in Fig. 2. One can see that with an increase in the pump rate, the maxima of the EM field at the boundaries of the active medium arise. Consequently, the number of harmonics building the collective mode also increases (see SM, Sec. 2). In other words, the number of states with the transition frequency increases forming a maximum of DOS.

Our computer simulation of the system with noise shows that the presence of an active medium also leads to a local increase in the DOS even if there is no reflection from the boundaries between the active medium and the outside space. In this case, the DOS changes due to the intermodal coupling through atoms. This coupling results in pulling free-space modes to the transition frequency of the inverted atoms and in the peak in the system DOS.

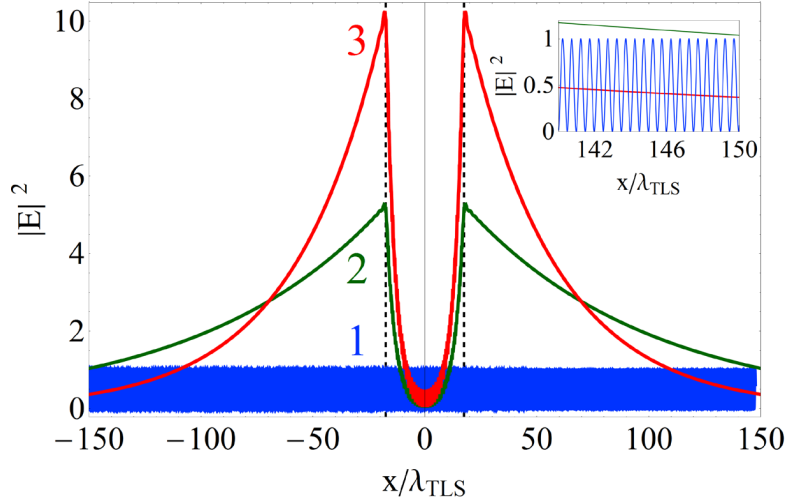

Fig. 2. The spatial distribution of the energy density of the electric field of the collective mode for different pump rates: $\gamma_P = \gamma_D$ (curve 1), $\gamma_P = 1.35\gamma_D$ (curve 2), and $\gamma_P = 1.48\gamma_D$ (curve 3). The inset shows field distributions at small scales.

Note, that the dynamics of mode pulling may only be observed in a system with noise [see Eqs. (2)-(4)]. Indeed, in such a system, the phase of the free-space mode fluctuates. Phase fluctuations may be presented as fluctuations in frequencies. Since at the transition frequency of active atoms the gain coefficient has a maximum, the closer to the transition frequency the mode frequency is, the slower is the mode decay. Therefore, the longest-living fluctuations are those that pull the mode towards the transition frequency. Such fluctuations give the greatest contribution to the mean frequency. To demonstrate this frequency pulling, we consider the spectrum of each space harmonic. For this purpose, we calculate the values of the correlator $A_j(\tau) = \langle a_j^*(t_{st} + \tau) a_j(t_{st}) \rangle / \langle a_j^*(t_{st}) a_j(t_{st}) \rangle$ and the spectrum of the $j$-th harmonic $S_j(\omega) = \mathrm{Re} \int_0^\infty d\tau A_j(\tau) \exp(i\omega\tau)$, in the stationary regime. Fig. 3 shows the frequency spectra of the harmonic with the wavevector $k_b = 0.978\omega_{TLS}/c$ for pump rates below, near, and above the threshold. Deep below the lasing threshold (the blue dashed line in Fig. 3), the spectrum has a pronounced maximum at $k_b = 0.978\omega_{TLS}/c$ and noticeable high-$k$ wing with a weak maximum at $k_{TLS} = \omega_{TLS}/c$. With increasing pumping, the first maximum decreases and the second maximum increases. Thus, the mean frequency of the space harmonics shifts towards the atomic transition frequency (Fig. 3, the green line). This maximum sharply grows with a further rate increase (the red dash-dotted line in Fig. 3). We refer to this phenomenon as the frequency pulling of space harmonics with different wavenumbers.

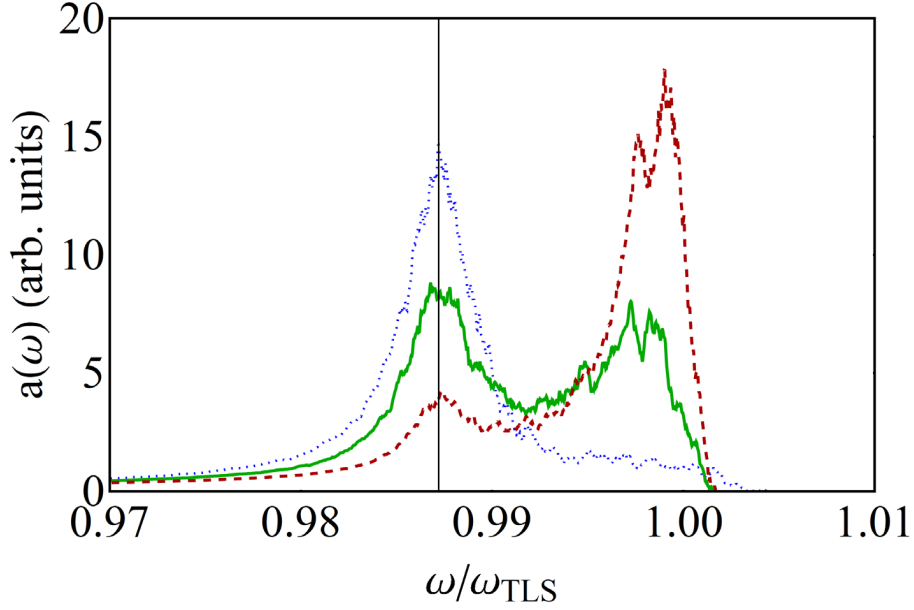

Fig. 3. Spectra of the free-space mode with the free-space eigenfrequency $\omega = 0.978\omega_{TLS}$ for different pump rates: $\gamma_P = \gamma_D$ (the dotted blue curve), $\gamma_P = 1.5\gamma_D$ (the solid thick green curve), and $\gamma_P = 2\gamma_D$ (the dashed red curve) for an extended system. Below the threshold, the maximum of the spectrum is at $\omega = 0.978\omega_{TLS}$ (marked by the vertical black line); with an increase in the pump rate, the maximum at the TLS transition frequency grows. $G(N_c)L_{am} = 28$.

Due to the frequency pulling, the collective mode, which consists of different space harmonics and is near the active medium, arises. Fig. 4 shows that near $k_{TLS} = \omega_{TLS}/c$, due to the frequency attraction of free-space modes to the transition frequency of active atoms, an interval of wavenumbers in which the dispersion curve tends to a horizontal line $\omega(k) = \omega_{TLS}$ arises. The size of this interval is determined by the level of noise and tends to infinity when noise vanishes. The flattening of the dispersion curve results in a decrease of the group velocity $v_{gr} = \partial\omega/\partial k$, which tends to zero, and a sharp increase in the DOS, which is inversely proportional to $v_{gr}$. Such a behavior of the DOS is shown in Fig. 5.

We find the DOS by calculating spectra $S_j(\omega)$ of harmonics. We assume that the value of $S_j(\omega)$ may be considered as a contribution of the $j$-th harmonic into the DOS dependence on the frequency, so that $\text{DOS}(\omega) = \sum_j S_j(\omega)$ [11]. Fig. 5 shows the DOS for pump rates near and far above the lasing threshold. One can see that the number of free-space modes increases in the frequency interval near the atomic transition frequency. For $\gamma_P = 1000\gamma_D$ it has a sharp maximum at this frequency. The maximum of the DOS and the zero of the group velocity lead to lasing in the system (see also Ref. [15]). Despite the absence of the resonator, this is similar to a conventional Fabry-Perot laser, in which the maximum of the DOS is created by a cavity [14]. In our case, the maximum of the DOS is the result of the nonlinear interaction between modes and atoms.

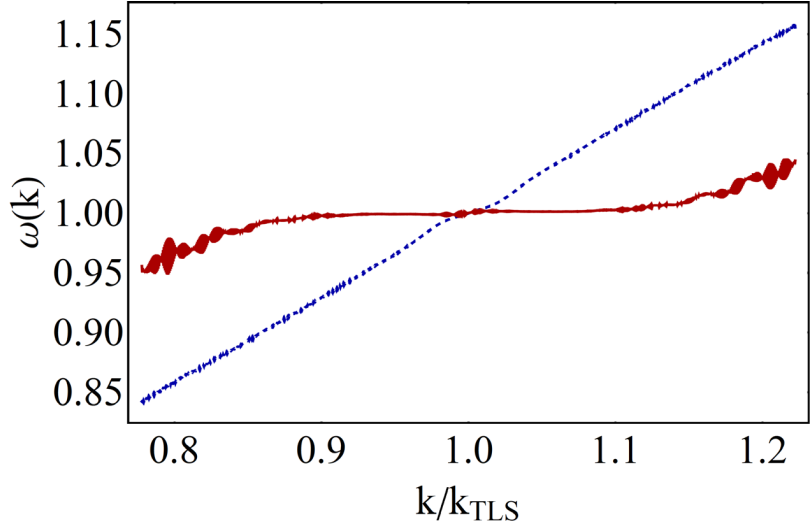

Fig. 4. The dependence of the mean frequency on the wavenumber for $\gamma_P = 2\gamma_D$ (the dashed blue curve) and $\gamma_P = 1000\gamma_D$ (the solid red curve).

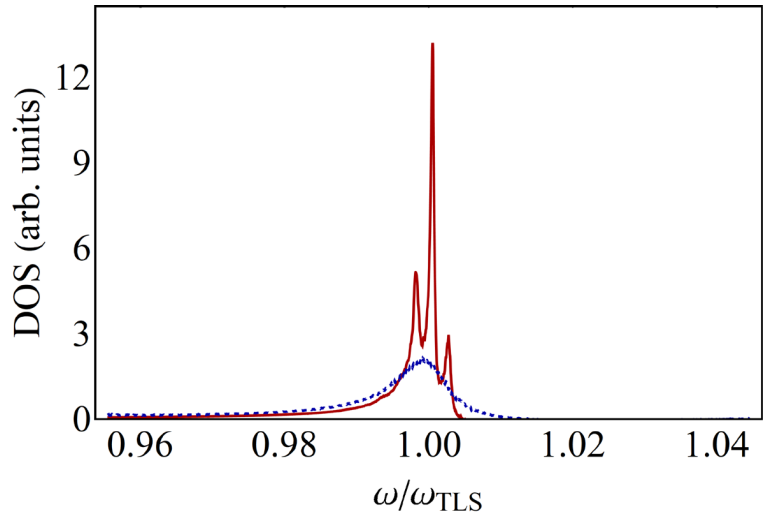

Fig. 5. The DOS near the transition frequency for $\gamma_P = 2\gamma_D$ (the dashed blue curve) and $\gamma_P = 1000\gamma_D$ (the solid red curve). The side maxima correspond to antisymmetric solutions, which have weaker interaction with atoms.

In Fig. 5, the maxima are caused by oscillations of the dispersion curve near the lasing threshold $\gamma_P = 1.48\gamma_D$. The coherence of the output radiation increases with the pump rate (see Fig. 1, the dash-dotted blue curve).

*Conclusion.* – We have demonstrated that above a coherence threshold, $\gamma_{coh}$, a cavity-free ASE system may generate coherent light. The reason for coherent radiation is the pulling of frequencies of free-space modes with different wavevectors towards the transition frequency of active atoms. This pulling occurs due to the nonlinear interaction of free-space modes with active atoms.

In a noisy system, we observe the mode pulling both below and above the coherence threshold. Moreover, the collective mode, which is formed in a long-lived fluctuation, arises at a pump rate far below the threshold. Such frequency pulling of the free-space EM field toward the transition frequency of active atom results in a peak in the DOS of the system and a sharp decrease in the group velocity of light. This decrease leads to an increase in the interaction strength between the EM field and the active medium [15], which ultimately leads to lasing.

In a system without noise, below the threshold, the mean value of the stationary field is zero. Nonetheless, a collective mode is being formed by the pulling of free-space harmonics. The eigenfrequency of this mode has a negative imaginary part. The real part of the eigenfrequency is almost equal to the atomic transition frequency. Above the coherence threshold, the imaginary part of the eigenfrequency becomes positive and the system radiates coherently. Thus, spontaneous emission is not essential to the formation of the coherently radiated ASE mode. As in a laser, spontaneous emission acts as a seed creating photons to initiate induced radiation. Indeed, in a system without noise, we find a sharp lasing threshold, as shown in Fig. 1.

Our analysis shows that the coherence threshold cannot be identified with the lasing threshold, which may arise due to reflection at the boundaries of an active medium and vacuum. To demonstrate this, we compare the DOS in our system with the DOS in the system with a Fabry-Perot resonator. The comparison shows that in order to obtain lasing at the coherence threshold due to feedback provided by the resonator, the amplitude reflection coefficient should be $r_{DOS} \approx 0.11$. This is much higher than the reflection from boundaries of an active medium and vacuum which in our system is $\sim 5.8 \cdot 10^{-3}$ (see Sec. 3 of the SM). In our system, the only condition necessary for lasing is a maximum in DOS.

*Acknowledgments.* – A.A.Z. thanks the Foundation for the Advancement of Theoretical Physics and Mathematics "Basis." A.P.V. and A.A.P. thank the 7-th program of the RAS for partial financial support.